\documentclass[twocolumn,english,aps,prb]{revtex4}
\usepackage[T1]{fontenc}
\usepackage[latin9]{inputenc}

\makeatletter

\providecommand{\tabularnewline}{\\}


\usepackage{babel}

\makeatother

\usepackage{babel}

\begin{document}

\title{The Pairing Symmetry in the Ferromagnetic Superconductor $UGe_{2}$}

\author{F. A. Garci\textbf{a, G. G. Cabrera }}

\affiliation{Instituto de Física {}``Gleb Wataghin'', UNICAMP, Campinas-SP,
13083-970, Brazil.}

\begin{abstract}
We give an extensive treatment of the pairing symmetry in the ferromagnetic
superconductor $UGe_{2}$. We show that one can draw important conclusions
concerning the superconducting state, considering only the transformation
properties of the pairing function, without assumptions about the
form of the pairing amplitudes.
\end{abstract}
\maketitle

\section{Introduction}

The interplay of superconductivity and magnetism is a subject of central
importance to accomplish a basic understanding of the fundamental
interactions in a solid. In particular, the coexistence of superconducting
and ferromagnetic states is still a matter of controversy. Early contributions
to this topic include the works of Matthias and Bozorth on $ZrZn_{2}$\cite{Matferro01}
and of Ginzburg \cite{Ginzspcf01}, who was the first to address the
problem of coexistence of both phenomena.

It was only recently that $UGe_{2}$ \cite{SaxenaUge} (for a more
complete reference see \cite{HuxleyUge}), was found to display such
properties. Since magnetic fields tend to suppress conventional superconductivity,
such a coexistence is expected to take place in a unconventional state
\cite{Manfred}.

Although one can find many efforts in the search for a microscopic
theory specific to this problem (\cite{Kirkpatrick,Zhou,Karchev}),
there is still a lot of work to be done, and a phenomenological approach
is of great importance.

The theory for unconventional superconductivity was given in \cite{Manfred}
for systems where the normal state (non superconducting) is paramagnetic,
and its group of symmetry includes the time reversal operation. In
turn, for ferromagnetic metals, the normal state is associated to
a magnetic group, or cogroup, and the superconducting states arising
from this normal state are classified according to the co-representations
of this cogroup \cite{Mineev}.

Consequences for $ZrZn_{2}$ were explored, including both, strong
and weak spin orbit couplings \cite{Samwsoc}. General discussions
on other systems were also given in \cite{Huxley}. Following the
above approaches, in this paper we propose a detailed and rigorous
treatment of this problem for the case of $UGe_{2}$, in the strong
spin orbit regime.

\section{The basic Hamiltonian and Symmetries.}

Our Hamiltonian is based on some aspects of the physics of $UGe_{2}$\cite{SaxenaUge,HuxleyUge}.
Its ferromagnetic (FM) properties resemble closely a metallic state,
where the Fermi surface has a big splitting ($70mev$) between the
$up$ and $down$ spin bands. This fact points for a choice of an
`equal spin pairing' ($ESP$) state, with intraband pairing only.
Moreover, the FM state is strongly anisotropic. The magnetization
is clearly pinned along the crystallographic $a$ axis, which we take
as the $z$ direction. This is very important, since the transverse
magnetic fluctuations are pair breaking for our proposed $ESP$ state.
Hence, these fluctuations are considered to be much smaller than the
interactions along the $z$ axis, and we keep just this component.

The basis of our theory is a generalized BCS Hamiltonian that allows
for triplet paring \cite{Ketterson}. We also add a term describing
a ferromagnetic exchange coupling with strength $J$ \cite{Zhou}.
After a standard mean field approximation, we remain with:

\[
\mathcal{H}_{MF}=\sum_{k\sigma}\epsilon_{k\sigma}c_{k\sigma}^{\dagger}c_{k\sigma}-\frac{1}{2}\sum_{k\sigma}(\Delta_{\sigma\sigma}^{*}(k)c_{-k\sigma}c_{k\sigma}\]
 \begin{equation}
+\Delta_{\sigma\sigma}(k)c_{k\sigma}^{\dagger}c_{-k\sigma}^{\dagger})-\frac{J}{2}\sum_{k}M(c_{k\downarrow}^{\dagger}c_{k\downarrow}-c_{k\uparrow}^{\dagger}c_{k\uparrow}),\label{eq: basicHamil}\end{equation}
with the definitions of the mean fields $M$ and $\Delta_{\sigma\sigma}(k)$
\begin{equation}
M=\frac{1}{2}\sum_{k^{\prime}}(\mathcal{h}c_{k^{\prime}\downarrow}^{\dagger}c_{k^{\prime}\downarrow}^{\dagger}\mathcal{i}-\mathcal{h}c_{k^{\prime}\uparrow}^{\dagger}c_{k^{\prime}\uparrow}^{\dagger}\mathcal{i}),\label{eq:defmag}\end{equation}
 \begin{equation}
\Delta_{\sigma\sigma}(k)=\sum_{k^{\prime}}V(k,k^{\prime})\mathcal{h}c_{-k^{\prime}\sigma}c_{k^{\prime}\sigma}\mathcal{i}.\label{eq:defgap}\end{equation}
 $M$ acts as a potential describing the magnetization of the system
and $\Delta_{\sigma\sigma}(k)$ is the pair potential of the superconducting
state. In our case, this potential is indeed a matrix that we write
in the form: \[
\hat{\Delta}(k)=\left(\begin{array}{cc}
\Delta(k)_{\uparrow\uparrow} & 0\\
0 & \Delta(k)_{\downarrow\downarrow}\end{array}\right),\]
which can be parametrized in terms of the elegant and useful $\vec{d}(k)$
vector formalism \cite{Ketterson}:

\begin{equation}
i(\vec{d}(k).\hat{\sigma})\hat{\sigma}_{y}=\left(\begin{array}{cc}
-d_{x}\left(k\right)+id_{y}(k) & d_{z}(k)\\
d_{z}(k) & d_{x}\left(k\right)+id_{y}(k)\end{array}\right).\label{eq:dvector}\end{equation}
With the choice of an $ESP$ state, we are left with a two-dimensional
order parameter lying on the $xy$ plane, while the spin of the Cooper
pairs is along the $z$ axis: \begin{eqnarray*}
d_{x}(k) & = & \frac{1}{2}(\Delta(k)_{\downarrow\downarrow}-\Delta(k)_{\uparrow\uparrow})\\
d_{y}(k) & = & \frac{i}{2}(\Delta(k)_{\uparrow\uparrow}+\Delta(k)_{\downarrow\downarrow}).\end{eqnarray*}
 For our purposes, a more suitable $\vec{d}(k)$ vector is: \begin{eqnarray}
d^{\Gamma}(k) & = & \frac{1}{2}[-(\hat{x}+i\hat{y})\Delta(k)_{\uparrow\uparrow}+(\hat{x}-i\hat{y})\Delta(k)_{\downarrow\downarrow}]\label{eq: d vector2}\end{eqnarray}
 where the $\Gamma$ index labels the irreducible corepresentation
associated with the transition. This $\vec{d}(k)$ vector is the order
parameter of our theory for triplet pairing. We denote by $G$ the
normal (ferromagnetic) state from which the superconducting state
will rise. It is written as: \begin{eqnarray}
G & = & M\times U(1)\ ,\label{eq: normalstate}\end{eqnarray}
 where $M$ is the magnetic group of the system, constructed from
its crystal symmetry, and $U(1)$ is the gauge group. $UGe_{2}$ has
an orthorhombic crystal structure with full inversion symmetry \cite{HuxleyUge},
identified with the $C_{mmm}$ space group. From now on, we treat
only the point group symmetry, as it is common when one consider the
Landau theory of phase transitions. It is based on the prescription
that macroscopic properties in thermodynamic equilibrium should not
depend on details related to the translational symmetry of the system.
The point group of interest is $D_{2h}$, which is generated by the
operations $\{C_{2z},C_{2x},I\}$. To describe a state with a magnetization
vector along the $z$ axis, one needs to include the time-inversion
operator $K$, combined with some symmetry operation. A suitable choice
that solves this problem is: \begin{equation}
\{E,C_{2z},KC_{2x},I\}\label{eq:maggroup}\end{equation}
 yielding the complete magnetic point group: \begin{equation}
M=\{E,C_{2z},KC_{2x},KC_{2y},I\}\ .\label{eq: pointgroup}\end{equation}
 Note that a magnetization (axial vector) along the $z$ axis is completely
invariant under the operations of (\ref{eq: pointgroup}), as it should
be. To further simplify our work, we only proceed with the group $\{E,C_{2z},KC_{2x},KC_{2x}\}$,
taking care in fixing the parity of the representations. This cogroup
is denoted $D_{2}(C_{2})$, and has two irreducible corepresentations,
which we label $A$ and $B$ (see table \ref{tab:Corepresen} ). %
\begin{table}

\caption{Corepresentations for the Cogroup $D_{2}(C_{2})$.}

\begin{centering}
\begin{tabular}{|c|c|c|c|c|}
\hline 
$\Gamma$  & $E$  & $C_{2z}$  & $KC_{2x}$  & $KC_{2y}$\tabularnewline
\hline
\hline 
$A$  & $1$  & $1$  & $1$  & $1$\tabularnewline
\hline 
$B$  & $1$  & $-1$  & $1$  & $-1$\tabularnewline
\hline
\end{tabular}\label{tab:Corepresen} 
\par\end{centering}
\end{table}

From table \ref{tab:Corepresen}, we see that $S_{A}$ is associated
with the trivial breaking of $U(1)$ gauge symmetry, while for $S_{B}$
we should consider a subgroup of $U(1)$, isomorphic to the invariant
subgroup $\{E,KC_{2x}\}$. Such a subgroup is simply $\{E,\exp(i\pi)\}$.
The result is that $S_{B}$ contains the so-called non-trivial elements.
Table \ref{tab:Allowed-Supercon} summarizes these results. %
\begin{table}

\caption{Allowed Superconducting Classes.}

\begin{centering}
\begin{tabular}{|c|c|}
\hline 
Classes  & Elements\tabularnewline
\hline
\hline 
$S_{A}(D_{2}(C_{2}))$  & $\{E,C_{2z},KC_{2x},KC_{2y}\}$\tabularnewline
\hline 
$S_{B}(\{E,KC_{2x}\})$  & $\{E,C_{2z}e^{i\pi},KC_{2x},KC_{2y}e^{i\pi}\}$\tabularnewline
\hline
\end{tabular}\label{tab:Allowed-Supercon} 
\par\end{centering}
\end{table}

\section{Analysis of Results}

Each of these superconducting classes is related to a gap structure.
This is determined from the transformation properties of the $\vec{d}(k)$
vector under the action of the class elements. For the $C_{2z}$ element
of the class $S_{A}$, and taking the $k$ vector as $k_{0}=(0,0,k)$,
we have: \begin{eqnarray*}
C_{2z}\vec{d}^{A}(k_{0}) & = & -\vec{d}^{A}(C_{2z}^{-1}k_{0})=-\vec{d}^{A}(k_{0}),\end{eqnarray*}
 but from table \ref{tab:Corepresen} we get: \begin{eqnarray*}
C_{2z}\vec{d}^{A}(k_{0}) & = & \vec{d}^{A}(k_{0})\ \ \Rightarrow\vec{d}^{A}(k_{0})=0\ ,\end{eqnarray*}
 meaning that the gap will have\emph{\ point nodes} on the intersections
of the Fermi surface with the $k_{z}$ axis. For the $S_{B}$ class,
taking $C_{2z}e^{i\pi}$ and the k-vector as $k_{0}=(k_{x},k_{y},0)$,
we have \[
C_{2z}e^{i\pi}\vec{d}^{B}(k_{0})=\vec{d}^{B}(C_{2z}^{-1}k_{0})=\vec{d}^{B}(-k_{0})\:.\]
 Recalling that $\Delta(k)$ must satisfy the Pauli antisymmetry property
$\Delta(-k)=-\Delta(k)$, and looking at table \ref{tab:Corepresen},
we write: \[
\vec{d}^{B}(k_{0})=C_{2z}e^{i\pi}d^{B}(k_{0})=(-1)d^{B}(k_{0})\ ,\]
 resulting that the gap will have a\emph{\ line node} on the intersection
of the Fermi surface with the $k_{x}k_{y}$-plane. Finite values for
the interband paring potential change this scenario. To investigate
its consequences, we directly study the transformation properties
of the paring potentials. First, we set up the transformation for
the $c_{k},c_{k}^{\dagger}$ operators under (\ref{eq: pointgroup}):
\begin{eqnarray}
(C_{2z}|\vec{0}):\lambda c_{k,\pm}^{\dagger} & \rightarrow & \mp i\lambda c_{C_{2z}k,\pm}^{\dagger}\nonumber \\
(KC_{2x}|\vec{0}):\lambda c_{k,\pm}^{\dagger} & \rightarrow & \pm i\lambda^{*}c_{-C_{2z}k,\pm}^{\dagger}\ ,\label{eq: opertransf}\end{eqnarray}
 where $\lambda$ is any complex number. To deduce these rules, one
should represent the time inversion as $K=(i\sigma_{y})\theta$ (where
$\theta$ is the complex conjugate operation), and consider that pseudo-spin
states transform like spin eigenstates. The translational part of
the operation is explicitly written, because we are now dealing with
properties of microscopic quantities in a way that lattice translations
are important. The situation is not more involved, since the space
group of interest is symorphyc. Now, we require invariance of the
Hamiltonian (\ref{eq: basicHamil}), including all the $\Delta_{\alpha\beta}$
potentials in the two body interaction part. Below, we work out an
example for the non-unitary operator $(KC_{2x}|\vec{0})$: \[
\sum_{k}[(KC_{2x}|\vec{0})\Delta_{\uparrow\uparrow}(k)c_{k\uparrow}^{\dagger}][(KC_{2x}|\vec{0})c_{-k\uparrow}^{\dagger}]=\]
 \[
=\sum_{k}\Delta_{\uparrow\uparrow}^{*}(k)[(-i)\ c_{-C_{2z}k\uparrow}^{\dagger}][(-i)\ c_{C_{2z}k\uparrow}^{\dagger}]\]
 \[
=\sum_{k^{\prime}}\Delta_{\uparrow\uparrow}^{*}(C_{2z}^{-1}k^{\prime})[c_{k^{\prime}\uparrow}^{\dagger}][c_{-k^{\prime}\uparrow}^{\dagger}]\]
 \[
\Rightarrow(KC_{2x}|\vec{0})\Delta_{\uparrow\uparrow}(k)=\Delta_{\uparrow\uparrow}^{*}(C_{2x}^{-1}k)\]
 We turn now to the expansion of the paring amplitudes in terms of
the cobasis functions $\Delta_{\Gamma}(k)=\sum_{i}\eta_{\Gamma,i}f_{i}(k)$.
The $\eta_{\Gamma,i}$ are the order parameter components usually
associated with the expansion of the Ginzburg-Landau free energy,
and $f_{i}(k)$ are the orbital basis functions. We choose the phase
as $KC_{2x}f_{\Gamma}(k)=f_{\Gamma}(k)$ for the basis functions,
being this choice dictated by the character table of this group. This
choice implies that we have $KC_{2x}\eta_{\Gamma}=\eta_{\Gamma}^{*}$.
We write \begin{eqnarray}
\Delta_{\sigma\sigma,A}(k) & = & \psi_{A}f_{B}(k)\nonumber \\
\Delta_{\sigma\sigma,B}(k) & = & \psi_{B}f_{A}(k)\ ,\label{eq: parintrans03}\end{eqnarray}
 with $\sigma=\uparrow,\downarrow$, and \begin{eqnarray}
\Delta_{\uparrow\downarrow,A}(k) & = & \psi_{A}f_{A}(k)\nonumber \\
\Delta_{\uparrow\downarrow,B}(k) & = & \psi_{B}f_{B}(k)\ .\label{eq: parintrans04}\end{eqnarray}
 Now, we can see that the symmetry dictated nodes for the interband
and intraband paring amplitudes are given by different corepresentations
. If one investigates, for example, the gap structure in the $(k_{x},k_{y})$
plane, even if the superconducting state happens to be in the $B$
state (realized trough $\Delta_{A}$), the above predicted nodal line
will be absent, since $\Delta_{A\uparrow\downarrow}$ relates to the
$f_{A}(k)$ amplitudes, which do not present this line. But, as already
discussed, we expect that a large splitting of the spin bands hinders
such interband components. Another interesting feature is that the
appearance of order parameters of the same symmetry in different sheets
of the Fermi surface, is connected with terms like $c_{k\uparrow}^{\dagger}c_{-k\uparrow}^{\dagger}c_{k^{\prime}\downarrow}^{\dagger}c_{-k^{\prime}\downarrow}^{\dagger}$
which do not conserve spin. This possibility is allowed due to the
strong spin orbit effect present in those systems.

To investigate further the gap structure, we consider specific basis
functions which come from the lattice Fourier series, thus taking
into account the translational orthorhombic symmetry \cite{Samwsoc}.
Considering odd functions, we have: \begin{equation}
f(k)=\sum_{n}c_{n}\sin(\stackrel{\rightarrow}{k}\cdot\stackrel{\rightarrow}{R}_{n})\label{eq: fourrie01}\end{equation}
 where $\stackrel{\rightarrow}{R}_{n}$ are the nearest neighbors
sites of the $UGe_{2}$ Bravais lattice \cite{HuxleyUge}. Using table
2 we get: \begin{eqnarray}
f(k)_{A} & = & \alpha\sin(\frac{ak_{z}}{2})\cos(\frac{ck_{y}}{2})\label{eq: basispart1}\\
f(k)_{B} & = & \beta\sin(\frac{ck_{y}}{2})\cos(\frac{ak_{z}}{2})\label{eq: basispart2}\end{eqnarray}
 Clearly, the basis (\ref{eq: basispart1}-\ref{eq: basispart2})
yields nodes other than the ones predicted before (`accidental nodes').
Note that all the symmetry dictated nodes are here, and we have just
determined extra nodes due to this possible particular choice of the
basis functions. Experiments should be used in order to distinguish
among the possible forms for the paring amplitudes. Theoretical predictions
(\cite{Samwsoc,WalkerUs}) may allow a meaningful interpretation of
experimental results. From those, ultrasound attenuation experiments
are most relevant for this kind of system. Such an experiment, acting
as a microscopic probe, would be very important in order to select
possible scenarios for the $f(k)$ functions. Considerations regarding
the structure dictated by a particular choice of basis functions are
important also, as it may act as a guide for the construction of a
microscopic theory. In principle, thermodynamic and transport properties
may distinguish between the $A$ and $B$ phases, since the topology
of the nodes implies different low temperature behaviors for such
quantities \cite{Manfred}. 

This work was supported by CNPQ and FAPESP (Brasil).


\begin{thebibliography}{10}
\bibitem{Matferro01} B. T. Matthias \& R. M. Bozorth. Phys Rev. 109,
604 (1957).

\bibitem{Ginzspcf01} V. L. Ginzburg. Soviet Physics JETP 4, 2 (1957).

\bibitem{SaxenaUge} S. S. Saxena et al. Nature 406, 10 (2000). 

\bibitem{HuxleyUge} A. Huxley et al. Phys Rev B 63, 144519 (2001). 

\bibitem{Manfred} M. Sigrist \& K. Ueda. Rev. Mod. Phys. 63, 2 (1991).
G. E. Volovik and L. P. Gorkov. Sov Phys JETP 61, 4 (1985). E. I.
Blount. Phys. Rev. B 32, 5 (1985).

\bibitem{Kirkpatrick} T. R. Kirkpatrick, D. Belitz. Phys. Rev. B
67, 024515 (2003).

\bibitem{Zhou} Y. Zhou, C. D. Gong. Europhys Lett, 74, 145-150 (2006).

\bibitem{Karchev} N. Karchev. Phys Rev B 67, 054416 (2003).

\bibitem{Mineev} V. P. Mineev. Phys Rev B 66, 134504 (2002).

\bibitem{Samwsoc} K. V. Samokhin and M. B. Walker. Phys Rev B 66,
024512 (2002); $ibid$ 174501 (2002).

\bibitem{Huxley} A. Huxley et al. Physica C 403, 9-14 (2004).

\bibitem{Ketterson} J. B. Ketterson \& S.N. Song. \emph{Superconductivity.}
Cambridge University Press (1999).

\bibitem{WalkerUs} M.B. Walker, M.F. Smith and K. V. Samokhin, Phys.
Rev. B 65, 014517 (2001). 
\end{thebibliography}
\end{document}